\documentclass[12pt]{article}
\usepackage{amsmath}
\usepackage{authblk}
\usepackage{graphicx}
\usepackage{color}
\usepackage{amsmath, amsthm, amssymb,amsfonts}
\usepackage{multirow}
\usepackage{natbib}
\usepackage[T1]{fontenc}
\usepackage{epic,eepic}
\usepackage[margin=2.5cm,dvipdfm]{geometry}
\title{Impact of a Liquid Drop on a Granular Medium:  inertia, viscosity and surface tension effects on the drop deformation}
\author[*,1]{E. ~Nefzaoui}
\author[]{O. ~Skurtys} 
\affil[1]{ Department of Food Science and Technology, Universidad de Santiago de Chile, Avenida Ecuador 3769, Santiago, Chile. Tel.: +56 2 718 4517}
\affil[*]{Current address: Institut P', CNRS-Universit\'{e} de Poitiers-ESIP, B\^{a}timent de m\'{e}canique, 40, Avenue du recteur Pineau, 86022 Poitiers Cedex, France}

\begin{document}
 \maketitle
\newpage
\begin{abstract} 

An experimental study of liquid drop impacts on a granular medium is proposed. Four fluids were used to vary physical properties: pure distilled water, water with glycerol at $2$ concentrations $1:1$ and $1:2$ v/v and water with Tween $20$ at the concentration of $0.1$g.l$^{-1}$. The drop free fall height was varied to obtain a Weber number ($We$) between $10$ and $2000$. Results showed that obtained crater morphologies highly depend on the impacting drop kinetic energy ($E_K$). Different behaviours during the drop spreading, receding and absorption are highlighted as function of the fluids viscosity and surface tension. Experimental absorption times are also commented and compared with a simplified theoretical model. Drops maximal extensions and craters diameters were found to scale as $We^{\frac{1}{5}}$ and $E_K^{\frac{1}{5}}$ respectively. In both cases, found dependencies are smaller than those reported in literature: $We^{\frac{1}{4}}$ for drop impacts on solid or granular surfaces and $E_K^{\frac{1}{4}}$ for spherical solid impacts on granular media.
\end{abstract}


\newpage
\section{Introduction}
The phenomenon of droplet impact on non-cohesive granular media is present in various industrial processes as food engineering, candy coated pills fabrication, starch wetting\dots \citep{Vaclavik}. It is also of great importance in environmental sciences: water erosion of soil, pollution of rivers and lakes, rainfall simulations \citep{vanDijk, Ries} \dots However, studies of this phenomenon are very scarce. \citep{Holman} studied the micro-metric drop impact on micro-metric grains of same sizes. Besides, authors used a very particular fluid, an aqueous polymer solution in this case. Others studied drops imbibition into powder beds with no impact. The drops were deposited on the granular surface with no kinetic energy \citep{Hapgood}. Very recently, the problem has been investigated by \citep{katsuragi} where the author only focused on the granular medium deformation, in particular the resulting craters morphologies for different drops velocities and granular media (grains sized between $4$ and $50\mu$m). Moreover, he reported the existence of different craters types depending on the drop free fall height and that water drop maximal extensions $D_{max}$, that he assumes equal to the crater diameter, scales as $We^{1/4}$. However, \citep{Nefzaoui} reported that, for water drops impact on coarser grains (average particle diameter $D_g \sim 81 \mu$m), $D_{max} \sim We^{1/5}$ and that the crater diameter $D_{rim}$ is not equal to  $D_{max}$.

On the other side, the drop impact phenomenon onto thin liquid films or dry solid surfaces has been thoroughly studied for over a century \citep{Yarin1}. The drop behaviour after collision with the surface is very complex since it depends on the physico-chemical characteristics of the drop and of the impact surface. The drop impact phenomenon can be divided in several sub-processes identified as spreading, receding, splashing and bouncing \citep{Rioboo, Vanderwal, Yarin1}. In recent years more attention has been shown to study the role of the surface roughness on the drop impact process \citep{Kannan, Xu}. In particular, it has been shown that the splashing of impacting drops is promoted by increasing the mean surface roughness.

Spherical solid impact into granular media were investigated to some extent decades ago \citep{Nelson, Caballero, Walsh, Zheng, Lohse}. For impact energies ranged between $10^{-2}$ and $0.6$ J, the morphology of the craters was reported in detail by \citep{Walsh}. In particular, the scaling of crater dimensions has been studied, and a power-law relationship between crater diameter $D_c$ and the energy of impact $E_K$ can be derived in certain limits: $D_c \sim E^{\frac{1}{4}}$. The process of crater formation is complex since the resistance of the granular medium to the projectile penetration depends on the medium intrinsic properties, its packing mode and the applied force. Indeed, impacting sphere penetration dynamics and grain ejection have been shown to be very different whether the granular material is loose or dense \citep{Lohse}. Another type of experiments concerns a bead impact on a granular target made of similar beads \citep{Beladjine}.
 
In this paper, an experimental study of drop impacts on a granular medium (glass beads) is proposed. Four different liquids were used in order to study the effect of viscosity and surface tension on the drop deformation for different impact velocities and try to determine the relevant parameters that governs the phenomenon. 
Different behaviours are highlighted depending on the receding magnitude and occurrence of splashing. The impact dynamics are also described through the spreading drop diameter temporal evolution obtained by high speed video. Maximal spreading drop diameters are compared to the case of drop impacts on solid surfaces \citep{Clanet, Rein}. Crater morphologies obtained by varying the drop kinetic energy at impact over more than three orders of magnitude are also presented and discussed. A power law for craters diameters dependence on the drop kinetic energy at impact is proposed and compared to solid projectiles impacts on granular media results.  
\section{Material and methods}
\subsection{Fluids and drop generation}
In our experiments, liquid drops were generated by a precision flat tipped syringe needle (Sigma-Aldrich, USA, St. Louis) connected to a digitally controlled syringe pump (Model 1000, New Era Pump System Inc., Farmingdale, NY, USA). Liquid flow rate was sufficiently low to get a nil drop initial velocity (0.05 ml.min$^{-1}$). Four fluids were used: pure distilled water, water with glycerol at 2 concentrations 1:1 and 1:2 v/v and water with Tween $20$ at the concentration of $0.1$g.l$^{-1}$. Their physical properties were measured and are presented in Table \ref{tab:fluids_phy_prop}. Three flat tipped needles (gauge $16$, $18$, $20$) were used to obtain various drop diameters $D_0$, between $2.88$ and $3.65$ mm. Drops are assumed with some assurance to maintain a spherical shape through the free fall since their radius were lower than the capillary length $\kappa$ defined by $\sqrt{\frac{\gamma}{\rho g}}$ \citep{Landau}. This was verified on the captured images for the biggest drops. Different impact velocities $V_0$ within the range of $0.4$ to $5$ m.s$^{-1}$ were obtained changing the drop release height $h$ up to $1300$mm. For each impact, $V_0$ was measured from captured images. The kinetic energy of the drop at impact, given by $E_K = \frac{ \pi}{6} \rho D_0^3 g h$ was varied between $1 \times 10^{-6}$ and $3 \times 10^{-4}$ J. Surface energy of the free drop surface, $E_S = \pi D_0^2 \sigma$ was varied between $1.8 \times 10^{-6}$ and $3 \times 10^{-6}$ J.
\subsection{Granular medium}
Target surface was composed of a circular container which was $78$ mm in diameter and $14$ mm deep. The container was filled with glass beads (G8893, Sigma-Aldrich, USA, St. Louis) which had an effective average particle diameter $D_g \sim 81 \mu$m and a density $\rho_g$ about $2300$ kg.m$^{-3}$. Granular medium compaction had a great effect on the crater formation. It was thus important to have the same initial beads bed state in general, and compaction in particular, for each drop impact. For this purpose, the same protocol was respected for each sample preparation. Beads were gently rolled into the container using a glass funnel. The container was overfilled and the surface was levelled using a straight-edge. There is no doubt that this operation introduced local compaction non-homogeneities. Finally, samples were weighed and the same mass of beads was used for each. Packing fractions of about $ 0.60 \pm 0.02 $ were deduced from the granular material measurements of the mass and volume of the bulk. 

\begin{table}
\begin{center}	
\small \begin{tabular}{@{}lcccccc@{}}
\hline 
Fluid & $\rho$ (kg.m$^{-3}$) & $\mu$ (mPa.s) & $\gamma$ (mJ.m$^{-2}$) & $D_0$ (mm) & $Oh$ $\times10^{3}$   \\
\hline
Water & 996  & 1  & 72.2 & 2.88 - 3.25 - 3.65 & $1.95$ - $2.2$\\
Water-glycerol 1:1 & 1121 & 6  & 66.9 & 3.47 & 11.8 \\
Water-glycerol 1:2 & 1162 & 19 & 65.1 & 3.39 & 37.5 \\
Water-tween $20$ ($0.1$g.l$^{-1}$) & 996  & 1  & 36 & 3 & 3 \\
\hline
\end{tabular}
\caption{Measured physical properties of fluids, the drop diameter ($D_0$) and the Ohnesorge number ($Oh=\frac{\mu}{\sqrt{\rho \gamma D_0}}$)}
\label{tab:fluids_phy_prop}
\end{center}
\end{table}

\subsection{Fluid-granular contact angle}
The contact angle of water over the granular media was measured through a capillary suction experiment as detailed by \citep{Xue}. The method consists in observing the velocity of a fluid rising in a granular column due to capillary forces. Taking into account hydrostatic, viscous and capillary  effects (only inertial effects are not considered here) leads to the following relation between the contact angle and the other physical parameters and variables \citep{Xue}:

\begin{eqnarray}
\gamma^2 cos^2(\theta) & = & \dfrac{2 \rho g \mu}{3} \dfrac{z^3 v}{z - 2tv}
\label{eq:washburn_gamma2_theta2}
\end{eqnarray}

where $\gamma$, $\rho$, $\mu$ and $\theta$ are the fluid surface tension, density, dynamic viscosity and the fluid-granular advancing contact angle respectively. Moreover, $t$ and $z$ are time and the rising fluid front position and $v$ is the fluid front velocity given by $v = \dfrac{dz}{dt}$. The only unknown is $\theta$ and can easily be deduced from Eq. (\ref{eq:washburn_gamma2_theta2}) since other parameters are measured experimentally.

\subsection{Granular medium pore diameter}
The granular medium mean pore diameter is needed to estimate the drop absorption times. According to \citep{Xue}, Washburn equation with hydrostatic effects leads to the following expression of the pore average radius $R_p$:

\begin{eqnarray}
R_p & = &  \dfrac{z^2}{\gamma cos(\theta)} \cdot \dfrac{1}{\dfrac{t}{2 \eta} - \dfrac{\rho g z^3}{3 \gamma^2 cos^2(\theta)}} 
\label{eq:wahsburn_viscous_gravitational_sol_dl}
\end{eqnarray}

This relation leads to consistent results when inertial effects are negligible, thus not at early times when $t$ and $z$ are too small.

\subsection{Images acquisition}
The impacting drop and the granular medium evolution, \textit{e.g.} the drop spreading and receding, crater formation, were observed experimentally with a high-speed video Pulnix TM-6740GE (Pulnix, Inc., San Jose, CA , USA) with a pixel resolution of $640 \times 480$. It could capture $200$ frames per second (fps) in full frame and $1250$fps with a $224 \times 160$ reduced matrix resolution. The camera was mounted on a boom stand which provided easy vertical or horizontal movement and stable support for the camera. A zoom video lens ($18-108$ mm f/$2.5$D, Edmund Optics, NJ, USA) was mounted on the video camera. Resulting frames were processed using ImageJ software (National Institutes of Health, USA).

\section{Results and discussion}
A mathematical formulation of the impact leads to consider the equation of continuity and the momentum equations both in the radial and in the axial directions. Three dimensionless numbers can be considered after the adimensionalization procedure: Reynolds ($Re=\frac{\rho V_0 D_0}{\mu}$), Weber ($We=\frac{\rho D V_0^2 }{\gamma}$), Froude ($Fr=\frac{V_0^2}{g D}$). In all tests, gravitational effect was negligible compared to inertia ($Fr^{-1} \ll 1$). Thus, only the Weber number, which compares the inertial forces to surface tension and the Reynolds number ($Re$), ratio of inertial to viscous forces were retained as parameters of the problem. Thus, the dynamics at impact was driven by an interplay between the kinetic energy, viscosity and the droplet surface tension. The adimensional number $\frac{We}{\sqrt{Re}}$ can also be used to take simultaneously into account the three parameters ($\frac{Inertia \cdot Viscosity}{SurfaceTension}$) \cite{fedorchenko2}. The three typical phases observed after the drop impact on the granular medium, \textit{i.e.} the spreading, receding and absorption, are detailed in the next paragraphs.

\subsection{Drop spreading dynamics}
\begin{figure}[hbt]
\centering
\includegraphics{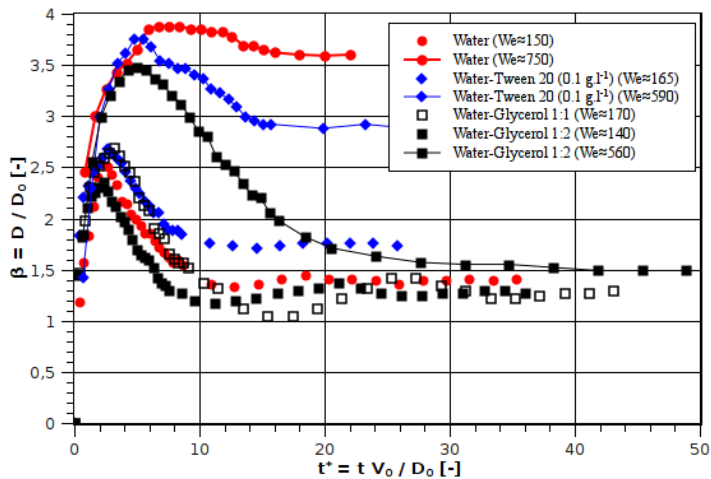}
\caption{Temporal evolution of the spreading factor $\beta = D / D_0$ for different Weber numbers and fluids.}
\label{beta_vs_tt0_all}
\end{figure}

\begin{figure}[hbt]
\centering
\includegraphics[scale=0.5]{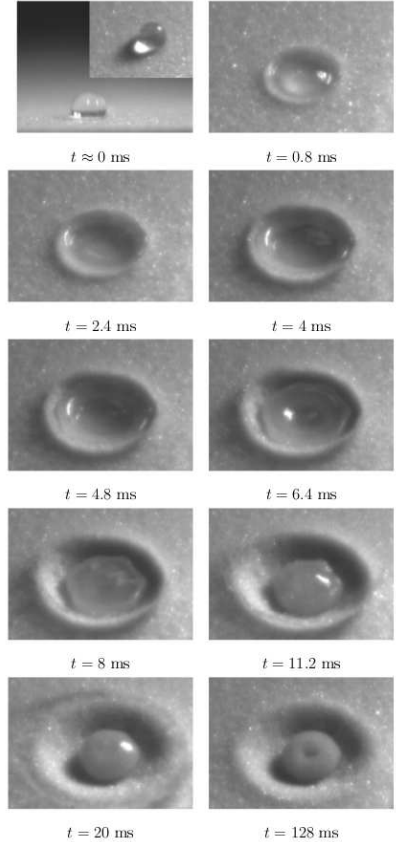}
\caption{Water drop impact images time sequence ($D_0=2.88$ mm; $We \sim 150$). Note: for $t \approx 0$ms, the top and front views are showed.}
\label{Sequence1}
\end{figure}

In Fig. \ref{beta_vs_tt0_all}, the temporal evolution of the spreading factor $\beta = D / D_0$ after impact is presented for high and low Weber numbers and different considered fluids. In a usual manner for inertia governed impacts, time $t$ is made nondimensional using the impact velocity $V_0$ and the initial spherical drop diameter $D_0$. In all cases, three phases can be observed: drop spreading ($\frac{d \beta}{dt} > 0$), receding ($\frac{d \beta}{dt} < 0$) and absorption. Spreading was the first stage of drop deformation after impact. The droplet initially flattens and spreads out horizontally into a pancake shape. Even if the description of the drop deformation before $t=0.8$ms was beyond our experimental capacities, it is tantalizing to say that in the first instants, the drop was not deformed and just penetrates into the granular medium like a spherical ball for early times as shown in Fig. \ref{Sequence1} (for $t \approx 0$). Indeed, in almost all first impact pictures, non-deformed spherical drops partially penetrating the granular medium were observed. This is not the case for a drop impacting on a solid surface where the drop deformation starts at impact time and its diameter evolution scales as $t^{\frac{1}{2}}$ \citep{Rioboo}. 

\begin{figure}[hbt]
\centering
\includegraphics[scale=0.5]{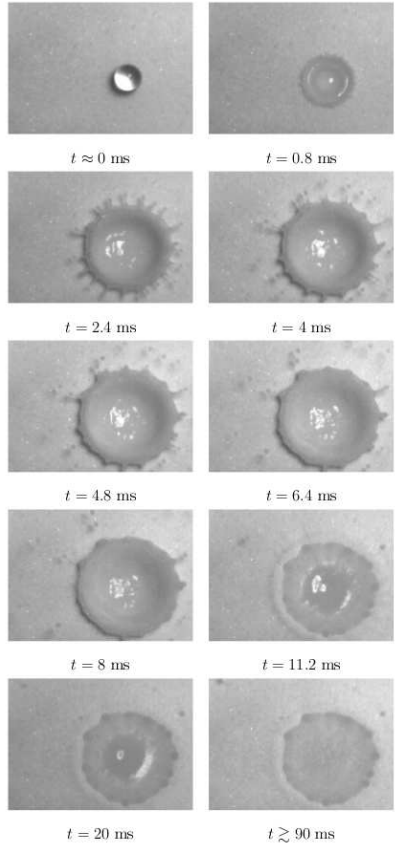}
\caption{Water drop impact images time sequence ($ D_0=3.65$ mm; $We \sim 650$).}
\label{Sequence2}
\end{figure}

Receding was the second stage of drop deformation after impact (see Fig. \ref{beta_vs_tt0_all}). In the case of water and Tween drops, the receding behaviour was considerably different at intermediate (and at low $We$ all the more) and high $We$. For $We \approx 150$, a "total receding" was observed, the drop final spreading factor $\beta_f$ is relatively small: $D_f \simeq 1.4 D_0$ (water) and $D_f \simeq 1.7 D_0$ (tween). Besides, when it reaches its final diameter, the fluid recovers a drop-like shape (see Fig. \ref{Sequence1}). Actually, at low and intermediate $We$, the drop spherical shape recovery let think that receding was governed by capillary forces. Whilst for higher kinetic energy, $We \approx 750$ (water) and $We \approx 590$ (tween), final drop spreading factor was closer to $\beta_{max}$: $D_f \simeq 3.5 D_0$ (water) and $D_f \simeq 2.9 D_0$ (tween). In contrast with the previous situation, we can talk about "partial receding" (see Fig. \ref{Sequence2}, for $t>8$ ms). In this case, the kinetic energy of the spreading fluid dominated the drop surface energy therefore the fluid groups into fingers along the outer rim of the droplet which had sufficient energy to overcome the surface tension and pinch off into smaller droplets. It is interesting to notice that a lower surface tension permits to increases surface area and hence $D_{max}$. In fact, for an impact at $We=590$, a Tween drop reached the same maximal extension than a water drop impacting at $We$ around $20\%$ higher ($We=590$) (see Fig. \ref{beta_vs_tt0_all}).
On the other hand, for highly viscous fluids (water-glycerol mixtures) with a surface tension close to water, the impacting drop receded to $\beta \leq 1$. Unlike water and Tween drops, no significative difference on the final drop diameter (after receding) was observed between high and low $We$ impacts (see Fig. \ref{beta_vs_tt0_all}). However, $D_{max}$ increases with $We$. Besides, oscillations of the spreading drop diameter were observed for low $We$. 

\begin{figure}[hbt]
\centering
\includegraphics{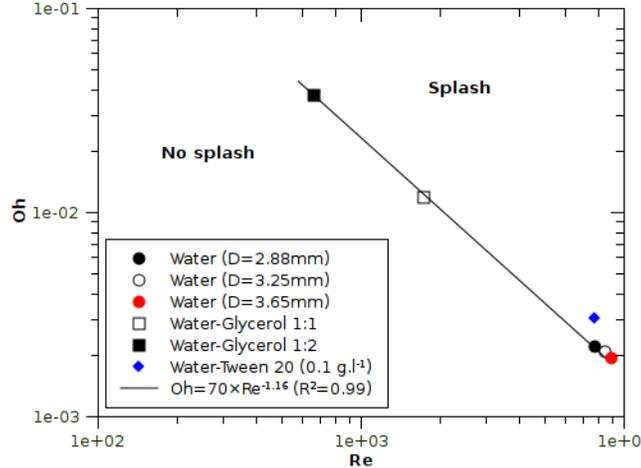}
\caption{Reynolds plotted versus Ohnesorge number showing the splash and no splash boundary.}
\label{splash}
\end{figure}

For the different considered fluids, splashing occurs at very high kinetic energy. To delineate the boundary between splashing and non-splashing regions a power-law relation based on $Oh$ and $Re$ numbers can be established: $Oh \times Re^{n} = b $ where $b$ is a real coefficient. Figure \ref{splash} shows $Re$ and $Oh$ values corresponding to the appearance of splashing. The boundary between splash-no splash then follows the relation: $Oh \times Re^{0.84} = 70$. This correlation can be compared with those reported by \citep{Kannan} for a drop impinging upon a dry solid surface or upon a thin fluid film where $Oh \times Re^{0.61} = 0.85$, $Oh \times Re^{1.17} = 63$ respectively. As $b_{film} \sim b_{granular}$, both exponents $n$ can be compared. To splash, the same drop impacting on a granular medium needs higher kinetic energy than those impacting upon a thin fluid film since $n_{film}>n_{granular}$.

\begin{figure}[hbt]
\centering
\includegraphics[scale=0.5]{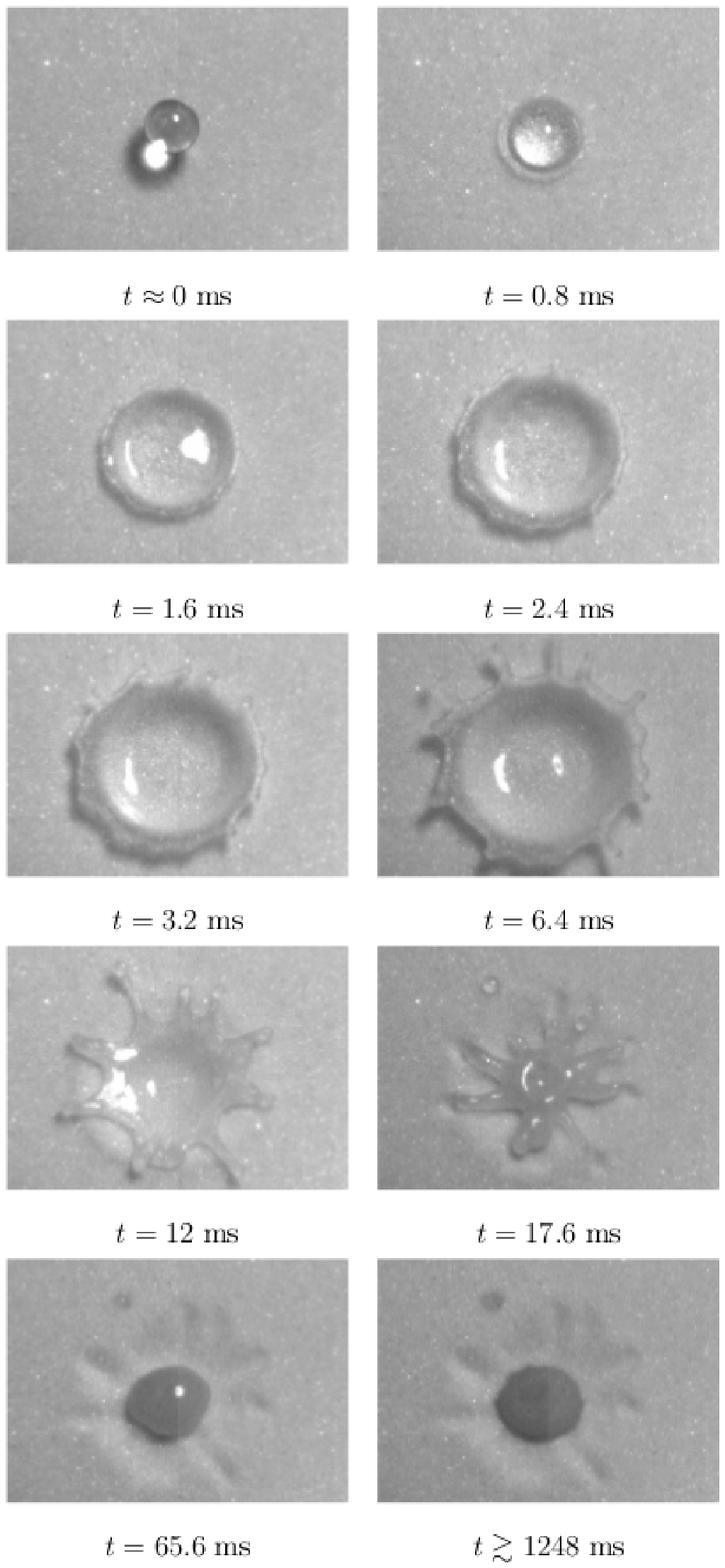}
\caption{Water-glycerol mixture (bulk ratio 1:1) drop impact images time sequence ($ D_0=3.47$ mm; $We \sim 570$).}
\label{Sequence3}
\end{figure}

\subsection{Maximum drop deformation}
\begin{figure}[hbt]
\centering
\includegraphics{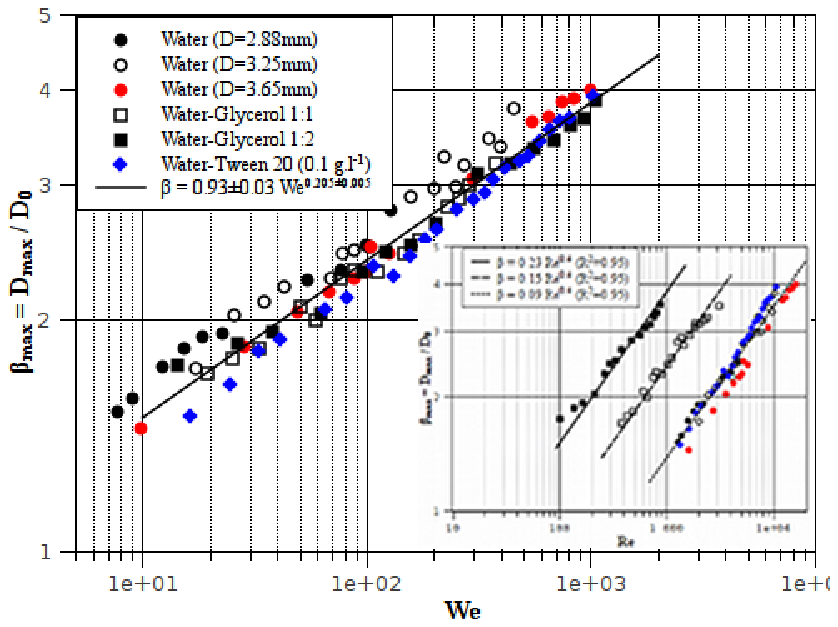}
\caption{The maximal spreading factor $\beta_{max} =D_{max}/D_0$ as function of the Weber number ($We$) for different drop diameters and fluids. Insert: The maximal spreading factor $\beta_{max} =D_{max}/D_0$ as function of the Reynolds number, $Re$.}
\label{Dmax=f(We)}
\end{figure}

In Fig. \ref{Dmax=f(We)}, the maximal spreading factor $\beta_{max} =D_{max}/D_0$ of drops impacting with different kinetic energies, is plotted versus $We$. All data were found to collapse in a $1/5$ power law: $\beta_{max} \propto We^{\frac{1}{5}}$ in spite of the surface tension and viscosity differences. A law of the type $\beta_{max} \propto We^{\frac{1}{5}}$ leads to $\beta_{max} \propto Re^{\frac{2}{5}}$ since $We^{\frac{1}{5}} \propto D_0^{-\frac{1}{5}}Re^{\frac{2}{5}}$. Our results do not concord with the models describing the drop impact on a solid surface reported by \citep{Clanet} and \citep{Rein}, $\beta_{max} \propto We^{\frac{1}{4}}$ and $\beta_{max}\propto Re^{\frac{1}{5}}$, respectively. Recently, \citep{katsuragi} reported that after the impact of a free-falling water drop onto a granular layer (where the grain size is varied between $4$ and $50 \mu$m), $\beta_{max}$ scales as $We^{\frac{1}{4}}$. In the case of a unique drop and even for drops of different diameters among values considered in this work ($(D_{01}/D_{02})^{\frac{1}{5}} \in [1,1.1]$), we can write $We^{\frac{1}{5}} \propto Re^{\frac{2}{5}}$. The insert of Fig. \ref{Dmax=f(We)} confirms our relation between the maximum spreading factor $\beta_{max}$ and the Reynolds number $Re$ for the four fluids with a high coefficient of determination ($R^2>0.95$).

\subsection{Drop absorption} 
Figures \ref{Sequence1} and \ref{Sequence2} show time sequences of a $2.88$mm and $3.65$mm diameter water drop impact for $We \sim 100$ and $We \sim 650$, respectively. Absorption is considered to be achieved when no visible difference between two and more consecutive frames can be observed. An absorption time of $128$ms was obtained in the first case and $90$ms in the second. It was surprising at first sight to find that the biggest drop took less time to be absorbed (an average bulk flow almost three times higher). After observing thoroughly the detailed frames sequences (which can not be shown here due to the high frames number), it has been found out that the mechanisms were not the same in the two cases. In the first case in particular, and for all impacts at $\dfrac{We}{\sqrt{Re}} \leqslant 1$ in general, the majority of drop absorption takes place after the drop spreading and receding and while the fluid is at rest. A fluid in a drop-like shape is then absorbed. Absorption and receding  are quite independent and the latter is governed by capillary forces. In the second case, absorption is observed to start relatively early and before the drop starts receding. This was observed for almost all impacts at $\dfrac{We}{\sqrt{Re}} \geqslant 1$ where a flat film is absorbed instead of a spherical drop. In fact, the higher the impact kinetic energy, the bigger the drop maximal extension and the thinner the spreading film. Besides, the fluid film absorption does not start at the periphery, where the film thickness may be thought to be the lowest. Actually, the fluid film thickness decreases from the center to periphery at first spreading stages \citep{Clanet}\citep{Roisman}. However, when the spreading lamella reaches its maximal extension, a blob appears at the rim. While there is no global fluid motion ($\frac{d \beta}{dt} = 0$ in Fig. \ref{beta_vs_tt0_all}), a local flow from the center to the periphery still exists, feeding the blob and thinning the fluid film at its center \citep{Roisman}. Fluid central sucking induces lamella receding. Therefore absorption is not the last stage of drop deformation. Absorption and receding are tightly coupled and we talk about an absorption induced receding. 
Table \ref{tab:abs_times} reports measured absorption times for different fluids and for different release heights. A few interesting remarks can be made.\\
First, these figures confirm that absorption times are systematically (except for very viscous fluids) higher for low release heights as explained above. Second, this difference of absorption times for different release heights decreases when the fluid viscosity increases.  Finally, it vanishes for very viscous fluids (Fig. \ref{Sequence3}). In fact, for glycerol-water 1:2 (19 times more viscous than water and water-tween mixture), no difference in absorption times is observed for different release heights while it can vary from one to 5 for water or water-tween mixture.\\
\begin{table}
\begin{center}
\begin{tabular}{lcccc}
\hline
Fluid & D(mm) & We & $t$(s) & $\tau$(s) \\
\hline
Water & 2.88 & 150 & 0.12 & 0.128 \\
Water & 3.65 & 750 & 0.09 & 0.09 \\
Water-tween & 3 & 164 & 0.27 & 1.25 \\
Water-tween & 3 & 549 & 0.05 & 1.25 \\
Water-Glycerol 1:1 & 3.47 & 114 & 1.68 & 6.3 \\
Water-Glycerol 1:1 & 3.47 & 570 & 1.06 & 6.3 \\
Water-Glycerol 1:2 & 3.39 & 119 & 2.29 & 14.1 \\
Water-Glycerol 1:2 & 3.39 & 546 & 2.43 & 14.1 \\
\hline
\end{tabular}
\end{center}
\caption{Measured ($t$) and calculated ($\tau$) absorption times for different drops of different sizes and fluids and falling from different heights.}
\label{tab:abs_times}
\end{table}

It can be interesting to calculate theoretically absorption times and to compare them with the experimental results. The method presented and confirmed in some particular cases by \citep{Hapgood} was adopted. In \citep{Hapgood} experiments, the drop was deposited with a syringe on the surface of the granular medium with no kinetic energy. With this method, the drop absorption was assumed to take place over a constant drop-granular medium contact surface (constant drawing area). Absorption time is then given by:

\begin{eqnarray}
\tau_{CDA} & = & 1.35 \dfrac{V_d^{\frac{2}{3}}}{\epsilon^2 R_p} \dfrac{\eta}{\gamma cos(\theta_d)}
\label{eq:tau_cda}
\end{eqnarray}
where $V_d$ is the initial drop volume, $\epsilon$ is the porous medium porosity, $R_p$ is the pore radius, $\gamma$, $\eta$ and $\theta_d$ are the fluid surface tension, viscosity and the fluid-granular medium contact angle respectively.

In our case, calculated times corresponded to experimental times in very few cases (See Table \ref{tab:abs_times}), for water drops falling from very low heights for example (experimental conditions very close to those of Hapgood) and no concluding results were obtained for water-tween and water-glycerol mixtures. The calculated times depended very highly on contact angles which were quite difficult to determine with a good precision. Besides, using this model implies the implicit assumption that no absorption occurs during spreading and receding which is obviously not true.
\subsection{Crater morphologies}
\begin{figure}
\centering
\includegraphics[scale=0.5]{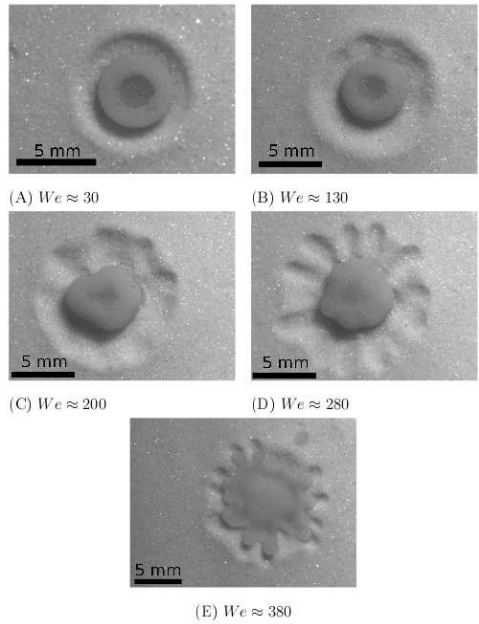}
\caption{Crater morphologies observed by dropping a $3.65$ mm diameter water drop into a container of glass beads for different Weber numbers.}
\label{Figure_impacts}
\end{figure}

In Figure \ref{Figure_impacts}, for different Weber numbers, examples of crater morphologies observed by impacting into a container of glass beads a $3.65$ mm diameter water drop are shown. The granular medium keeps a memory of the fluid shape at its maximal extension, thanks to its deformability and wettability. Similar craters were also observed with the different fluids. At low $We$, \textit{e.g.} $We \approx 30$ (Fig. \ref{Figure_impacts}-A), a crater with in its centre a circular agglomeration of glass beads was formed. A rotationally symmetric rim was raised above the original target surface level. Glass beads driven by the receding drop formed a central agglomeration where the surface energy is sufficiently strong to join glass beads by capillary cohesion forces. This crater morphology was observed for $We \lesssim 100$. As $We$ was increased the first irregularities appeared inside the crater and the agglomeration in the centre was less circular (Fig. \ref{Figure_impacts}-B). These irregularities became more marked at higher $We$ (Fig. \ref{Figure_impacts}-C-D). They can be associated with the fingering instability at the rim of the expanding water drop like reported in the previous studies of the water drop on a solid surface \citep{ALLEN75, MEHDIZADEH04}. This crater morphology occurred in the range $100 \lesssim We \lesssim 300$. This kind of crater was not reported by \citep{katsuragi}. In its work, finer grains were used and lower $We$ numbers were considered which may explain these differences. At higher kinetic energy, a layer of wet glass beads appears instead of the central agglomeration previously described (Fig. \ref{Figure_impacts}-E). It is due to the water drops splashing.

\begin{figure}
   \centering
   \includegraphics{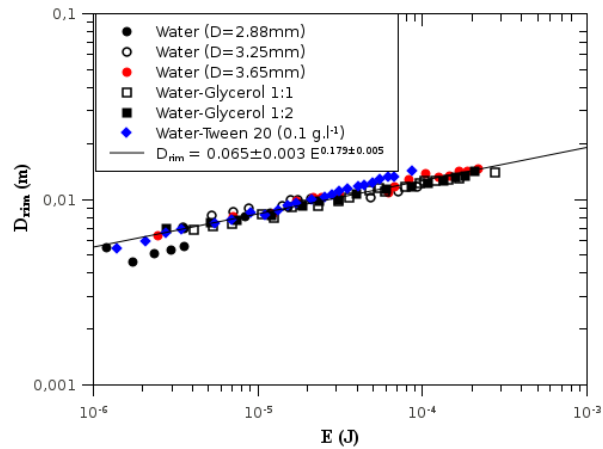}
   \caption{Crater rim diameter ($D_{rim}$) as function of the kinetic energy ($E_K$) for different drop diameters and fluids.}
   \label{D=f(E)}
\end{figure}

In Figure \ref{D=f(E)}, the crater rim diameter $D_{rim}$, measured at the top of the crater rim, is plotted against kinetic energy of drops at impact. As expected, for different drop diameters and fluids, the measured diameter $D_{rim}$ increased with increasing $E_K$. Moreover, the data are well described by a $\frac{1}{5}$ power law over more than three orders of magnitude in $E_K$. However, this law is slightly different of those reported for the impact of a solid projectile where $D_{rim}$ scales as $E_K^{\frac{1}{4}}$ \citep{Walsh,Uehara03} and the water drop impact reported by \citep{katsuragi}. 

\section{Conclusion}
An experimental study of liquid drop impacts on a granular medium has been presented for Weber numbers ($We$) in the range $10-2000$. Four liquids were used to vary either the viscosity (between $1$ and $19$mPa.s) or the surface tension (between $36$ and $72$mJ.m$^{-2}$). Different crater morphologies have been observed depending on the Weber number. The crater rim diameter was found to scale as a $1/5$ power-law of the kinetic energy of drops at impact $E_K$. This law was slightly different of those reported for the impact of a solid projectile where $D_{rim}$ scales as $E_K^{\frac{1}{4}}$. 

The temporal evolution of the spreading factor ($\beta = D / D_0$) was presented. The effects of viscosity, surface tension and inertia on the spreading, receding and absorption were commented. A power-law $\beta_{max} = D_{max} / D_0 \sim We^{\frac{1}{5}}$ smaller than those recently reported by \citet{katsuragi} has been deduced for the different fluids. 

The boundary between splash-no splash was found to follow the relation: $Oh \times Re^{0.84} = 70$. In particular, to splash, the same drop impacting on a granular medium needs higher kinetic energy than those impacting upon a thin fluid film.

Measured absorption times depended highly on $We$ and the fluid viscosity. Calculated times according to the simplified model reported by \citet{Hapgood} were not in agreement with experiments. A better prediction of absorption times must be achieved taking into account absorption during spreading and receding.

Finally, more imagery acquisition is necessary to understand the fluid granular interactions (crater formation mechanisms) and to develop an energetic model that details the initial drop kinetic energy dissipation during the drop deformation (viscous dissipation, \dots) and crater formation (granular dissipation).  

\section*{Acknowledgements} 
The authors thank F. Melo for discussions and help, the CONICYT sponsor of the Bicentenario Grant (PBCT PSD-$62$) and Dicyt (project $081071SO$).

\newpage 
 
\bibliographystyle{plainnat}
\bibliography{drop}

\end{document}